\documentstyle[preprint,tighten,prl,aps]{revtex}
\begin{document} \draft
\title{Impurities in S=1/2 Heisenberg Antiferromagnetic Chains:
Consequences for Neutron Scattering and Knight Shift}
\author{Sebastian Eggert$^a$, Ian Affleck$^{b}$}
\address{$^a$ Theoretical Physics, Chalmers University of Technology,
 and G\"oteborg University, 41296 Gothenburg, Sweden and
$^{b}$Canadian Institute for Advanced Research and Physics Department
 University of British Columbia,  Vancouver, B.C.,V6T 1Z1, Canada}
  \date{\today} \maketitle
\begin{abstract}
Non-magnetic impurities in an S=1/2 Heisenberg antiferromagnetic chain
are studied using boundary conformal field theory techniques and
finite-temperature quantum Monte Carlo simulations. We calculate the
static structure function, $S_{\rm{imp}}(k)$, measured in neutron
scattering  and the local susceptibility, $\chi_i$ measured in Knight
shift experiments.    $S_{\rm{imp}}(k)$ becomes quite large near the
antiferromagnetic wave-vector, and exhibits much stronger temperature
dependence than the bulk structure function. $\chi_i$ has a large
component which alternates and {\it increases} as a function of distance from
the impurity. \end{abstract}
\pacs{75.10.Jm,75.20.Hr}
Although the spin chain problem has been a popular topic for
theoretical physicists since the early days of quantum mechanics,
the correlation functions of the antiferromagnetic
Heisenberg spin-1/2 chain could
only be calculated with the help of modern quantum field
theory\cite{peschel}.  Adding non-magnetic impurities to a spin-chain
compound breaks the chains up into finite sections with essentially
free boundary conditions. The correlation functions in the presence of
such a boundary were calculated only rather recently \cite{eggert}.
These results provide a simple application of a general theory of
conformally invariant boundary conditions which has been applied to a
wide variety of quantum impurity problems in condensed matter and
particle physics\cite{affleck1}.  These functions exhibit a universal
dependence on the boundary, at long distances and times.  In this
paper we wish to focus on a couple of applications of these results of
experimental relevance:  the impurity contribution to the static
structure function,  $S_{\rm{imp}}(k,T)$, and the local
susceptibility, $\chi_i(T)$.  We derive analytic expressions for these
quantities using field theory methods and compare them with finite-T
Monte Carlo simulations using lengths of up to $l=128$ with  a varying
number of time-steps up to 64   and several hundred thousand sweeps
through each lattice.

The Heisenberg Hamiltonian for the antiferromagnetic spin-1/2  chain
\begin{equation} H = J\sum_i \vec S_i\cdot \vec S_{i+1} \label{ham}
\end{equation} is equivalent to a free boson field theory in (1+1)
dimensions in the low energy, long-distance limit\cite{affleck2}.
The spin operators are expressed in terms of the boson $\phi$ as
\begin{equation} S^z_j \approx  \partial_x \phi/\sqrt{2\pi}
 + a (-1)^j  \cos{\sqrt{2\pi}\phi },
\label{cont-op}\end{equation} where $a$ is a constant. The
boson Hamiltonian is then simply given by the free part together with
terms which become irrelevant as the temperature is lowered.  Those
irrelevant terms give rise to temperature and finite length dependent
corrections with a characteristic power-law.

This theory has been used successfully to calculate impurity
effects\cite{eggert}, the low energy spectrum\cite{schulz},  and
correlation functions\cite{peschel}. The latter agree well with recent
neutron-scattering experiments\cite{Tennant}. Like the expression for the spin
operators in equation (\ref{cont-op}), the correlation functions also acquire
an alternating and a uniform part as a function of the site index $x$.  At
finite-temperature, the alternating part is given by:
\begin{equation}<S^z(x,t_1)S^z(y,t_2)>_{\rm{alt}}\to c  {\pi\over v\beta}
{(-1)^{x-y}\over\sqrt{\sinh {\pi (x-y-v\Delta t)\over v\beta}
 \sinh {\pi (x-y+v\Delta t)\over v\beta}}}
\label{finite-TN}  \end{equation}
($\Delta t \equiv t_2-t_1$.)  We set the lattice spacing to 1. The spin-wave
velocity is known to be $v = J \pi/2$  from the Bethe-ansatz. The constant
$c$ can be determined numerically and  is given by $c = a^2/2$ times an
arbitrary normalization of the two-point function, which we chose to set to
one. The irrelevant terms in the
Hamiltonian give logarithmic corrections to this expression\cite{schulz}.
In fact, it has been shown recently that the logarithmic corrections
give rise to an infinite slope of the uniform susceptibility
at zero temperature\cite{suscept}.

The correlation functions in the presence of a boundary were
first calculated in reference \onlinecite{eggert}.  There it was
argued that the free boundary condition on the spin operators
corresponds, in the continuum limit, to a boundary condition on the
bosons: $\phi (0)=\phi_L(0)+\phi_R(0)=\sqrt{\pi /8}$.
 Since $\phi_L$ is a function only of $vt+x$ and
$\phi_R$ of $vt-x$, this implies that we may simply regard this boundary
condition as defining $\phi_R$ to be the analytic continuation of $\phi_L$ to
the negative axis $\phi_R(x)=-\phi_L(-x)+\sqrt{\pi /8}$.
 Whereas the bulk correlation function factorizes into a product of left
and right 2-point Green's functions, the boundary correlation function
becomes a 4-point Green's function for left-movers.  Consequently, while
the uniform part is largely unaffected by an open boundary condition, the
alternating part gets modified to \begin{equation}
 c \ (-1)^{x-y}\frac{\pi}{v \beta}\sqrt{\frac{\sinh{2 \pi x\over
v \beta} \ \sinh{2 \pi y\over v \beta}}
{\sinh{\pi (x+y+ v \Delta t)\over v\beta}
\sinh{\pi (x+y- v \Delta t)\over v\beta}
\sinh{\pi (x-y+ v \Delta t)\over v\beta}
\sinh{\pi (x-y- v \Delta t)\over v\beta}}} \label{finite-T}
\end{equation}
which reduces to equation (\ref{finite-TN}) in the bulk limit
$x y \gg |(x-y)^2 - v^2 \Delta t^2|$.

Here we have also included the time-dependence of the Green's function, but we
will only calculate the equal-time spatial Fourier transform, $S(k)$,
deferring consideration of the full dynamical structure function to later
work. We predict a characteristic impurity contribution
to the structure factor, which may be observable in magnetic
Neutron Scattering experiments on quasi one-dimensional spin-1/2
magnetic compounds (e.g. KCuF$_3$).
Doping with impurities will break the spin-chains and thereby introduce
the desired open ends.  For a finite chain of length $l$ we can define
a structure factor $S_l(k)$ as
\begin{equation}
S_l(k) \ \equiv \ \frac{1}{l} \sum_{x,y=1}^l <S^z(x) S^z(y)> e^{ik(x-y)}
\ \  \stackrel{l\to \infty}{\longrightarrow} \ \
S(k) + {S_{\rm{imp}(k)}\over l}. \label{Sk}
\end{equation}
The structure function for the finite chains
has been decomposed into a ``bulk'' part $S(k)$
which is independent of length and an ``impurity'' part of order $1/l$:
\begin{equation}
S_{\rm imp}(k) \equiv \lim_{l \to \infty}l [S_l(k) - S(k)].
\end{equation}
The bulk part
reproduces the signal of a system without open ends (e.g.
an infinite chain) while the effect of the open boundary condition
is entirely contained in the  impurity part.  Higher
order ${\cal O}(1/l^2)$ terms
will also be present, but can be neglected if the impurities are  dilute.
Since each impurity creates the same contribution $S_{\rm imp}(k)$
in the dilute limit, the experimental signal will contain the
impurity part as a term which scales with impurity concentration
$n$ to first order:
\begin{equation}
S_{\rm exp}(k) \ \approx \ S(k) + n S_{\rm imp}(k).
\end{equation}

{}From the results of equations (\ref{finite-TN}) and (\ref{finite-T}), it is
clear that we expect interesting effects for wave-vectors
near $k \approx \pi$.  Field theory predictions for small
$k-\pi$ and $T$ are obtained by Fourier transforming equations
(\ref{finite-TN}) and (\ref{finite-T}).
We assume here that the impurities are dilute enough so that the
infrared cut-off is always given by the inverse temperature $\beta \ll l/v$.
The bulk structure
function\cite{Schulz2} can then be expressed in terms of the digamma
function $\psi$\cite{AS} and the reduced variable $k' \equiv
(k-\pi)v \beta/\pi$:
\begin{equation} S(k') = 2c\left[ \ln (\Lambda\beta J)- {\rm Re}\
\psi \left(1/2
-ik'/2\right) \right],\label{S(k)ft}\end{equation} where
$\Lambda$ is a  constant depending on the cut-off.

The impurity contribution, $S_{\rm{imp}}(k')$ is obtained by Fourier
transforming equation (\ref{finite-T}) with the bulk part, from equation
(\ref{finite-TN}), subtracted off assuming two open ends.
This subtraction eliminates the
ultra-violet divergence, giving the scaling form: \begin{eqnarray}
S_{\rm{imp}}(k') & = & c  { 2 v\beta \over
\pi}\int_0^{\infty}dw\left[\int_{0}^w du{\cos {k' u}
\over \sinh u}\left(\sqrt{1- {\sinh^2 u \over \sinh^2
 w}}-1\right)
 -\int_w^\infty du{\cos {k' u}\over \sinh u}
\right] \nonumber \\ & = & v\beta f(k') \label{Simp(k)ft} \end{eqnarray}
Here $u=\pi (x-y)/ v\beta$, $w=\pi (x+y)/ v\beta$.
Note that, apart from the logarithmic term in equation (\ref{S(k)ft}), $S(k')$
and $S_{\rm{imp}}(k')$ are functions only of the scaling variable
$k'=(k-\pi)v \beta/\pi$,
but we expect corrections to this scaling behavior from irrelevant operators
and the ultraviolet cut-off which become smaller as $T\to 0$ and
$k-\pi \to 0$ with $k'$ held fixed.
For small $k'$ we have  $S(k')  \propto \ln(\beta) +
{\rm const.} + {\cal O}(k'^2)$ and  $S_{\rm imp}(k') \propto v\beta
[{\rm const.} + {\cal O}(k'^2)]$, so that the impurity part has a much
stronger temperature dependence. The second term in equation (\ref{Simp(k)ft})
can be written $\int_0^{\infty}dw  \ \int_{w}^\infty du
\ {\cos {k' u} \over \sinh u} = - {d \over dk'} {\rm Im} \psi(1/2-ik'/2)$,
which is the dominant contribution for small $k'$.
At large $k'$, however, the leading behavior comes from the first integral
$S_{\rm imp}(k')\to - v \beta c/ \pi k'^2$, while $S(k')$ vanishes
exponentially.  (In the opposite limit $l/v \ll \beta \to \infty$
we get a delta-function at $k=\pi$
for both the impurity and the bulk contributions
$S_{\rm imp}(k=\pi) \propto l$ and $S(k=\pi) \propto \ln(l)$.)

\begin{figure}
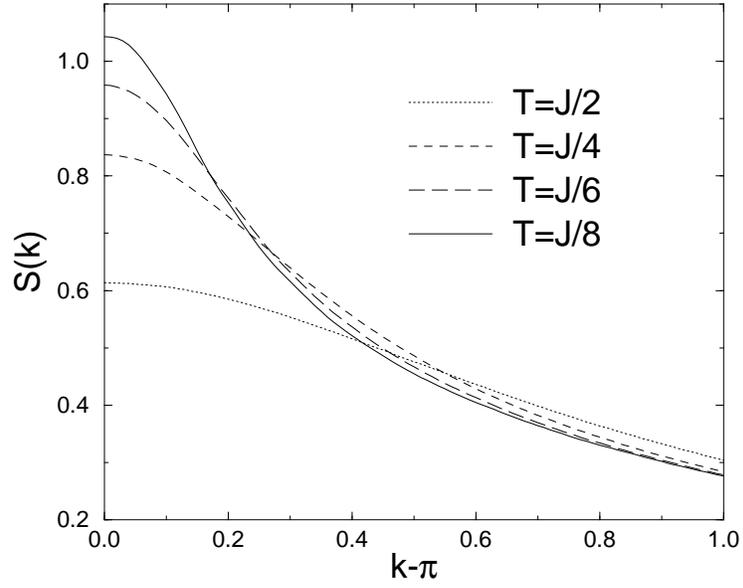

\caption{The bulk
structure factor $S(k)$
according to quantum Monte Carlo simulations.} \label{Sk-mc}
\end{figure}

\begin{figure}
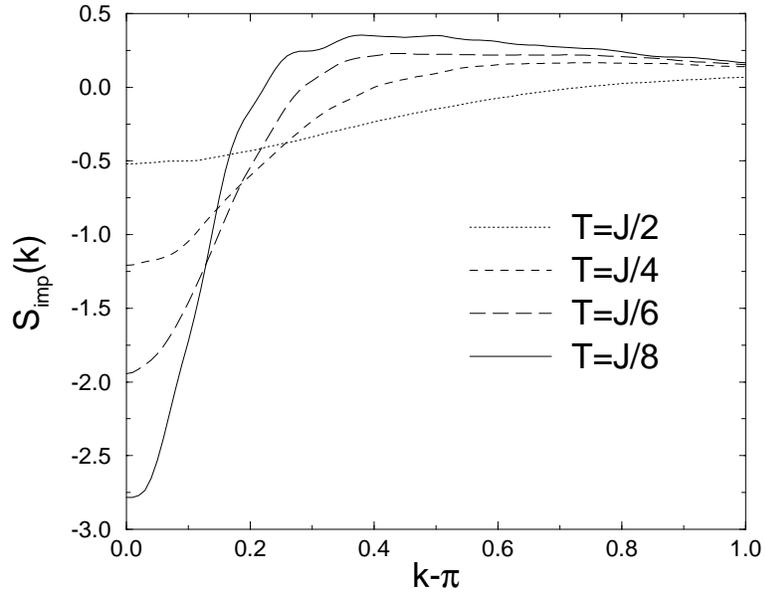

\caption{The impurity part of the structure factor $S_{\rm imp}(k)$
according to  quantum Monte Carlo simulations.} \label{Sk-imp-mc}
\end{figure}

\begin{figure}
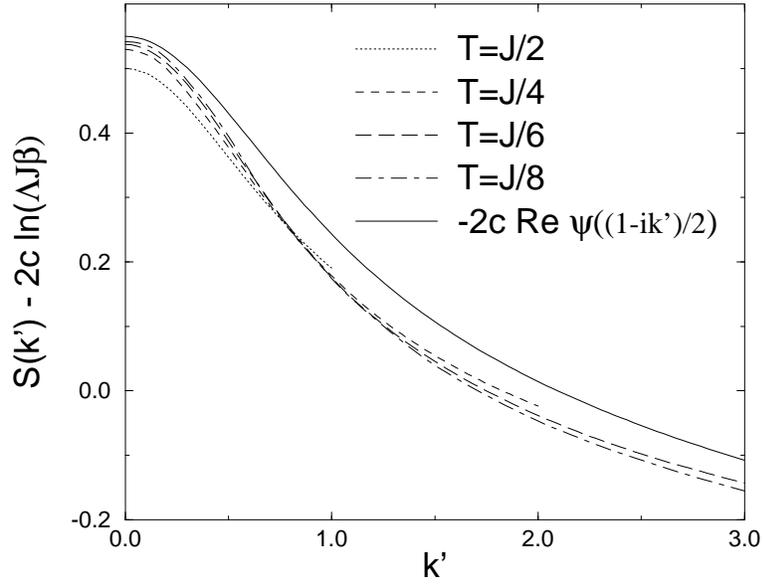

\caption{Monte Carlo results for the shifted bulk structure
function, $S(k')-2c\ln (\Lambda J\beta )$
compared to the field theory prediction
of equation (\protect{\ref{S(k)ft}}),
 with $c=0.14$ and $\Lambda = 0.75$.}\label{Sk-comp} \end{figure}

\begin{figure}
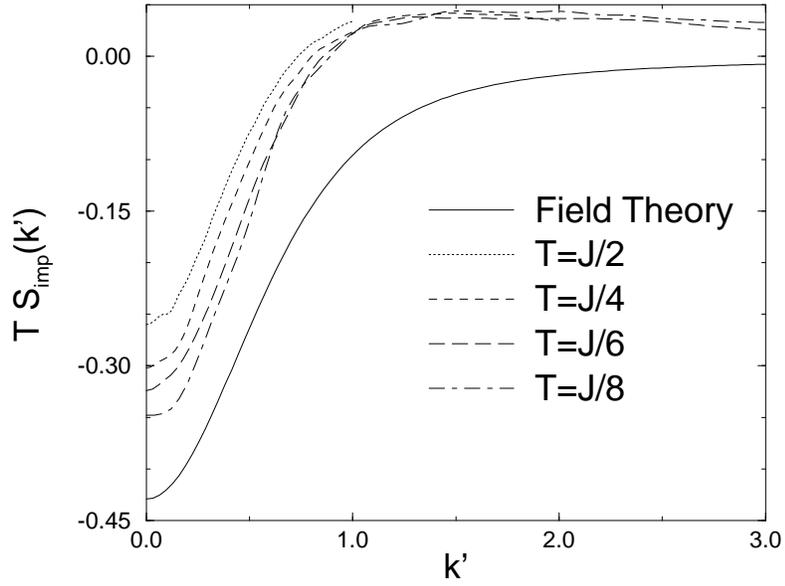

\caption{Monte Carlo results
for the scaled impurity part $TS_{\rm imp}(k')$
compared to equation
 (\protect{\ref{Simp(k)ft}}) with $c=0.14$.}
\label{Sk-imp-comp} \end{figure}

Our Monte Carlo results for $S(k)$ and $S_{\rm{imp}}(k)$ are
shown in figures (\ref{Sk-mc}) and (\ref{Sk-imp-mc}) for four temperatures.
Note, that since the impurity part scales with the inverse temperature
$\beta$, it may make a sizable contribution even at moderate impurity
densities.  To show the predicted scaling form  we plotted  the results
as a function of the reduced variable $k'$
in figures (\ref{Sk-comp}) and (\ref{Sk-imp-comp}).
The Monte Carlo  simulations
agree reasonably well with the field theory predictions.
  This comparison with the Monte Carlo data was used to
extract the constant  $c = 0.14$ and $\Lambda=0.75$
in equation (\ref{S(k)ft}).

We now consider the local susceptibility $\chi_i$ at any arbitrary site $i$
under the influence of a uniform magnetic field $h$ acting on the
complete chain
\begin{equation}
\chi_i (T)  \ \equiv  \ \frac{\partial}{\partial h} <S^z_i>|_{h=0}  =
{1\over T}\sum_j< S^z_jS^z_i>, \label{chi_i} \end{equation}
For a chain with periodic boundary conditions, $\chi_i$ is the same
for all sites because of translational invariance.

If we are dealing with an open boundary condition, however, the
translational invariance is clearly broken and we would naively
expect the open end to be more susceptible.  Moreover, it is
now possible, in the field theory treatment, to have a non-zero alternating
susceptibility as a function of site
index $\chi_x = \chi_x^{\rm uni} + (-1)^x \chi_x^{\rm alt}$.
 Using the analytic continuation of the
left-movers onto the negative half axis from above,
$\chi_x^{\rm alt}$
is given by a non-zero three point Green's function:
\begin{eqnarray}
\chi_x^{\rm alt} & \equiv   &
\beta <  S^z_{\rm alt}(x)  \int dy S^z_{\rm uni}(y) >
\nonumber \\ & = &     \frac{a \beta}{\sqrt{8 \pi}}
 \int_{-\infty}^{\infty} dy  \left< ie^{-i\sqrt{2 \pi}
\phi_L(x,t^\prime)} e^{i\sqrt{2 \pi}
\phi_L(-x,t^\prime)} \frac{\partial \phi_L}{\partial x}(y,t)
\ + \ h.c. \right>  \nonumber \\
& = &  \frac{a \beta}{4 \pi} \int_{-\infty}^{\infty} dy
\frac{\sqrt{{v\beta \over \pi}
\sinh {2\pi  x\over v\beta} }}{\frac{v \beta}{\pi}  \sinh {\pi \over
v \beta} (y+x+iv\Delta \tau) \ \ \frac{v \beta}{\pi}
\sinh {\pi \over  v \beta} (y-x+iv\Delta \tau)}
\nonumber \\ & = & \frac{a}{v} {x \over \sqrt{\frac{v \beta}{\pi}
\sinh {2 \pi x \over v \beta}}}
 \ \stackrel{\beta\to \infty}{\longrightarrow} \ \
{a \over v}\sqrt{x\over 2},
\label{alt-susc}
\end{eqnarray}
where $x$ is the distance from an open boundary condition.
  At low temperatures  the alternating part actually {\it increases} with
the distance $\sqrt{x}$ from the open end.  Any finite temperature
suppresses this growth  exponentially with $x,$  so that we expect a
typical maximum which gets
shifted further into the chain as the temperature is lowered.
Furthermore, even at $T=0$, the staggered magnetization does not increase
indefinitely with distance from the impurity, but rather oscillates  with a
wavelength, $4\pi v/h$,
i.e. \begin{equation}
M^{\rm{alt}}(x,h,T=0) = a\sqrt{2\over x}\sin ({hx\over 2v}).
\end{equation}
This exotic behavior is similar to Friedel oscillations except that
the $1/r^3$ decay which occurs there gets enhanced to a $\sqrt{r}$
{\it growth} due to a combination of reduced dimensionality and the absence
of charge fluctuations in this pure spin system.

\begin{figure}
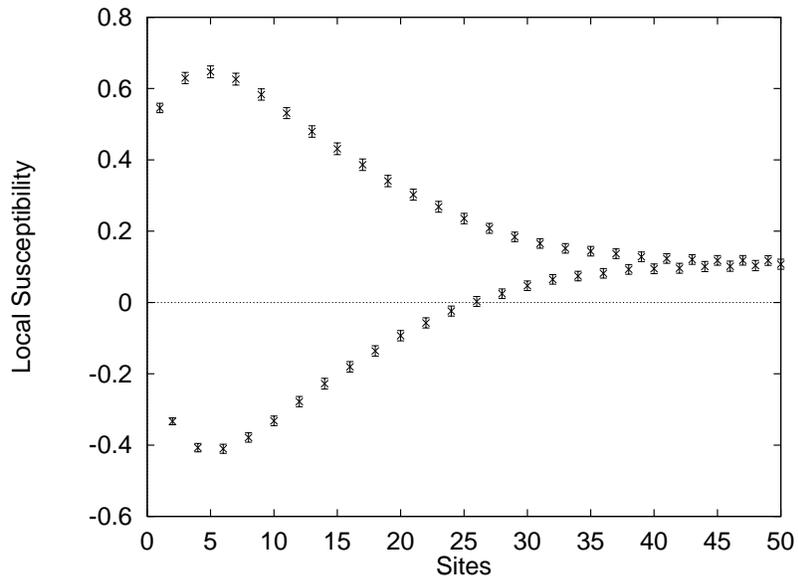

\caption{The local susceptibility vs. distance
from the  open end according to Monte Carlo
simulations at $T = J/15$.}
\label{susc1} \end{figure}

\begin{figure}
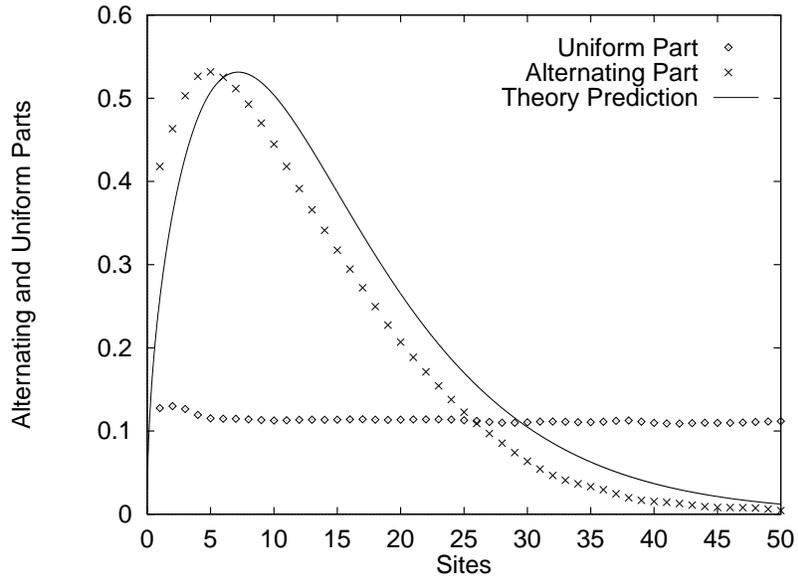

\caption{The uniform and alternating parts of the local susceptibility
according to Monte Carlo
simulations at $T = J/15$ compared to the field theory
equation (\protect{\ref{alt-susc}})
with $a=0.58$.} \label{susc2}
\end{figure}

The result from equation (\ref{alt-susc}) can be
confirmed independently with quantum Monte Carlo simulations.
The local susceptibility as a function of distance from the open end
is shown in
figure (\ref{susc1}) from Monte Carlo simulations at $T = J/15$.
After extracting the uniform and alternating parts as shown in figure
(\ref{susc2}), we can compare the alternating part to the predicted form
from  equation (\ref{alt-susc}) where the overall constant was chosen to
be $a = 0.58$.  The field theory prediction $c=a^2/2$, together with the
value $c=0.14$ from our MC measurement of $S(k)$ gives $a\approx 0.53$ in
reasonable agreement. While the shape of the theoretical prediction for
$\chi_x^{\rm{alt}}$ fits the Monte Carlo results very well, there is an
unexplained shift of about two sites, which might be due to irrelevant
operators. The functional dependence in equation (\ref{alt-susc}) holds
rather well for all temperatures $\beta$ sampled (up to the shift of two
sites).  For $T=J/15$ the shift in the
susceptibility due to the impurity is larger than the bulk susceptibility
over a distance
 of about 25 lattice sites from the impurity.  Thus we expect that
it should be possible to observe this effect in nuclear magnetic resonance
Knight shift experiments.  Note, that
$\chi_i <0$ for small even $i$, so that
those spins will tend to anti-align with the applied field.

The uniform part of the susceptibility is not directly
affected by the boundary condition, but gets an additional
non-universal contribution near $x=0$ from an irrelevant boundary
operator\cite{eggert,thesis}, which also appears
to be present in the Monte Carlo results in figure (\ref{susc2}).
This shift in the uniform susceptibility is what would be expected
classically, but the large alternating part is a purely quantum
mechanical effect.

In conclusion, we have calculated the effect of impurities on the
neutron-scattering cross-section and the NMR Knight shift using both field
theory and Monte Carlo methods.  The two methods are in reasonable
agreement and the effects seem large enough to be  observable
experimentally. The Knight shift actually {\it increases}
with distance from the impurity in the limit of zero field and temperature.
\begin{center}ACKNOWLEDGEMENTS \end{center}

We would like to thank Bill Buyers, Junwu Gan, Henrik Johannesson,
Rob Kiefl, Erik S\o rensen
and Eugene Wong for helpful discussions.  This research was supported in
part by NSERC of Canada and the Swedish Natural Science Research Council.

 

\begin{references} \bibitem{peschel} A. Luther, I. Peschel, Phys. Rev.
\underbar{B12}, 3908 (1975). \bibitem{eggert}S.~Eggert, I.~Affleck, Phys.\
Rev.\ \underbar{B46}, 10866, (1992).
\bibitem{affleck1} I. Affleck, {\it Correlation Effects in Low-Dimensional
Electron Systems} (ed. A. Okiji, N. Kawakami, Springer-Verlag, Berlin
1994), p. 82.
\bibitem{affleck2}For a review of the conformal field theory treatment of
the spin-1/2 chain and earlier references see I. Affleck, {\it Fields,
Strings and Critical Phenomena}  (ed. E. Br\'ezin, J. Zinn-Justin
North-Holland, Amsterdam 1990), p.563.
\bibitem{schulz} I. Affleck, D.
Gepner, H.J. Schulz, T. Ziman, J. Phys.  \underbar{A22}, 511, (1989).
\bibitem{Tennant} D.A. Tennant, T.G. Perring, R.A. Cowley,
S.E. Nagler, Phys. Rev. Lett. \underbar{70}, 4003 (1993). \bibitem{suscept} S.
Eggert, I. Affleck, M. Takahashi, Phys. Rev. Lett. \underbar{73}, 332 (1994).
\bibitem{thesis}  S. Eggert, ``Impurity Effects in Antiferromagnetic
Quantum Spin-1/2 Chains'', Ph.D. Thesis, University of British Columbia,
Vancouver, August 1994.
\bibitem{Schulz2}H.J. Schulz, Phys. Rev. \underbar{B34}, 6372 (1986).
\bibitem{AS} M. Abramowitz and I. A. Stegun {\it Handbook of Mathematical
Functions} (Dover, New York, 1964).
\end{references}
\end{document}